\documentclass[preprint]{aa}
\usepackage{psfig}
\usepackage{epsfig}
\usepackage{natbib}
\bibpunct{(}{)}{;}{a}{}{,} 
\def \sax{BeppoSAX}
\def \U {4U\thinspace1812--12}

\def\ergscm{ergs s$^{-1}$ cm$^{-2}$}
\def\ergs{ergs s$^{-1}$}
\def \nh {N${\rm _H}$}

\def \hcm {\hbox {\ifmmode $ atom cm$^{-2}\else atom cm$^{-2}$\fi}}
\def \arcmin {\hbox{$^\prime$}}

\def\comptt{{\em Comptt}}
\def\rxte{RXTE}
\def\sax{BeppoSAX}
\def \chisq {$\chi ^{2}$}

\def\approxgt{\mathrel{\hbox{\rlap{\lower.55ex \hbox {$\sim$}}
        \kern-.3em \raise.4ex \hbox{$>$}}}}
\def\approxlt{\mathrel{\hbox{\rlap{\lower.55ex \hbox {$\sim$}}
        \kern-.3em \raise.4ex \hbox{$<$}}}}

\begin{document}


\title{Simultaneous BeppoSAX and Rossi X-ray Timing Explorer observations of \U}

\author{D. Barret\inst{1}, J. F. Olive\inst{1}
        \and T. Oosterbroek\inst{2}  
}
\offprints{Didier Barret (Didier.Barret@cesr.fr)}

\institute{
Centre d'Etude Spatiale des Rayonnements, CNRS/UPS, 9 Avenue du
       Colonel Roche, 31028 Toulouse Cedex 04, France
\and
ESA-ESTEC,  Science Operations \& Data Systems Division,
SCI-SDG, Keplerlaan 1, 2201 AZ Noordwijk, The Netherlands
}

\date{Received / Accepted}

\abstract{\U~is a faint persistent and weakly variable neutron star X-ray binary. It was observed by \sax\ between April 20th and 21st, 2000 in a hard spectral state with a bolometric luminosity of $\sim 2\times10^{36}$\ergs. Its broad band energy spectrum is characterized by the presence of a hard X-ray tail extending above $\sim 100$ keV. It can be represented as the sum of a dominant hard Comptonized component (electron temperature of $\sim 36$ keV and optical depth $\sim 3$) and a weak soft component. The latter component which can be fitted with a blackbody of $\sim 0.6$ keV and equivalent radius of $\sim 2$ km is likely to originate from the neutron star surface. We also report on the first measurement of the power density spectrum of the source rapid X-ray variability, as recorded during a simultaneous snapshot observation performed by the Rossi X-ray Timing Explorer. As expected for a  neutron star system in such hard spectral state, its power density spectrum is characterized by the presence of a $\sim 0.7$ Hz low frequency quasi-periodic oscillation together with three broad noise components, one of which extends  above $\sim 200$ Hz. 
}
\titlerunning{BeppoSAX/RXTE observations of \U}
\authorrunning{Barret, Olive \& Oosterbroek}
\maketitle

\section{Introduction}
\label{sect:intro}
\U~(also known as Ser X-2) was discovered with Uhuru by \citet{forman76}, while the first X-ray bursts from this source were detected with Hakucho by \citet{murakami83}. This established a neutron star as the compact object  in \U, and the system to be a likely low-mass X-ray binary (LMXB). 

During the monitoring of the Galactic Center region with the Wide  Field Cameras of \sax, several X-ray bursts were detected from  \U. In most of these bursts, clear photospheric radius  expansion due to Eddington-limited burst luminosity was detected  allowing to estimate the source distance to 4 kpc \citep{cocchi00}.

Very little is known about the continuum source spectrum. EXOSAT detected the source as part of its Galactic Plane Survey \citep{warwick88}. \citet{gottwald95} using the EXOSAT GSPC instrument reported a featureless power law spectrum in X-rays (below 20 keV). Although these observations suggested that \U\ could possibly emit hard X-rays, the source has never been detected above ~20 keV.

Here we report on the first detection of \U~in the hard X-ray range with the \sax~X-ray satellite, and the first measurement of its rapid X-ray variability with the Rossi X-ray timing explorer (RXTE). 
\section{Observations and results}
\begin{figure}[!t]
  \centerline{\includegraphics[width=7cm]{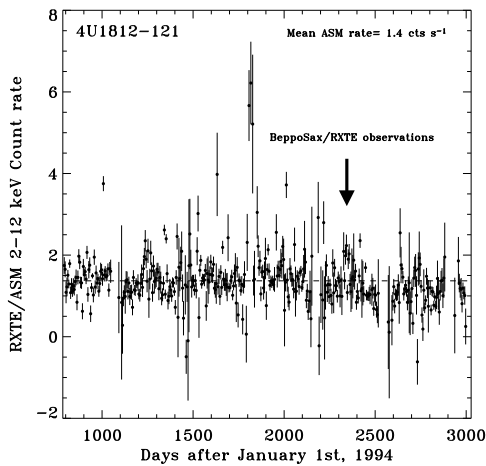}}
  \caption[]{The RXTE/ASM 2-12 keV light curve of \U~(each data point is a time average over 5 days). The time of the \rxte/\sax~pointed observations is indicated with an arrow. The mean source count rate is equivalent to about 20 mCrab in X-rays. The raw ASM data dowloaded from HEASARC have been filtered using the FTOOLs {\it fselect} to satisfy the criteria that the reduced \chisq~of the fit to recover the source intensity is less than 1.5. For information about the ASM, see \citet{levine96}.}
  \label{barret_fig1}
\end{figure}

RXTE/All-Sky Monitor (ASM) data indicate that \U~is a faint, persistent and weakly variable X-ray source at the level of $\sim 20$ mCrab in the 2-12 keV range. This is illustrated in Fig. \ref{barret_fig1} which represents the RXTE/ASM light curve over more than 6 years. We observed \U~between April 20th and 21st, 2000 with both \sax~and \rxte. The time of these observations is indicated with an arrow on Figure \ref{barret_fig1}. The ASM data recorded during the \sax~observation indicated that the source was slightly brighter than its average level (1.7 counts s$^{-1}$ against 1.4 counts s$^{-1}$, see Fig. \ref{barret_fig1}). In the next section, we present the spectral results obtained from \sax~and present the timing results from the simultaneous \rxte~observation.

\subsection{\sax}
\label{sect:obs}
Results from the Low-Energy Concentrator Spectrometer  \citep[LECS; 0.1--10 keV,][]{parmar97lecs}, the Medium-Energy Concentrator Spectrometer \citep[MECS; 1.8--10 keV,][]{boella97mecs} and the Phoswich Detection System \citep[PDS; 15--300 keV,][]{frontera97pds} on-board \sax\ are presented. All these instruments are coaligned and collectively referred to as the Narrow Field Instruments, or NFI. Unfortunately,  no High Pressure Gas Scintillation Proportional Counter data was obtained, since the instrument was switched off during the observation.

The region of the sky containing \U\ was observed by \sax\ between 2000 April 20 (18:51 UT) to April 21 (13:23 UT). Good data were selected from intervals when the instrument configuration was nominal, using the SAXDAS 2.0.0 data analysis package. LECS and MECS light curves, images and spectra were extracted using radii of 8\arcmin\ and 4\arcmin, respectively. Background subtraction for the imaging instruments was performed using standard files, but is not critical for such a bright source. Background subtraction for the PDS used data obtained during intervals when the collimators were offset from the source. The LECS and MECS spectra were rebinned to oversample the full width half maximum of the energy resolution by a factor of 3 and to have at least 20 counts per bin to allow use of the $\chi^2$ statistic. A LECS response matrix appropriate for the position and count rates of the source was generated.  The PDS spectrum was rebinned using the standard techniques in SAXDAS. The LECS, MECS, and PDS exposure times are 15.5 ks, 32.7 ks, 14.4 kiloseconds. 

A visual inspection of the light curves indicated that no X-ray bursts were detected during the observation. The image analysis revealed that \U\ was detected at a position of R.A.=18h 15' 06.4" and Dec=-12$^\circ$ 05' 27.1" (J2000 coordinates). This position is fully consistent with the more accurate one derived from the ROSAT all sky survey \citep{voges99}, during which \U\ was detected as a bright point source at a position R.A.=18h 15' 06.1" and Dec= -12$^\circ$ 05' 45.0" (ROSAT total positioning error of 10").

For the spectral analysis, we selected data in the energy ranges 0.5--10.0~keV (LECS), 1.8--10~keV (MECS), and 15--200~keV (PDS) where the instrument responses are well determined and sufficient counts obtained. Due to the high \nh\ we have not used data below 0.5 keV.  A systematic uncertainty of 1\% has been added to the data. Factors were included in the spectral fitting to take into account the small differences in the absolute flux determination of the various NFI instruments. These factors were constrained to be within their usual ranges during the fitting\footnote{For information about the cross-calibration of the \sax~instruments, see www.asdc.asi.it/bepposax/software/cookbook/cross\_cal.html.}. The spectral fitting and error computations were carried out using XSPEC 11.2 \citep{arnaud96}.

The overall spectrum of \U~was investigated by simultaneously fitting data from all the \sax\ NFI. In X-rays, the spectrum of \U\ is strongly absorbed. This was accounted for by using a photoelectric absorption model based on the cross sections of \citet{morrison83}. \U\ was clearly detected in the PDS up to $\sim 100$ keV. The hard X-ray tail was fitted with a Comptonization model. In the present paper, we used the \comptt\ model, whose description is given in \citet{titarchuk94}. \comptt\ is an analytical model which describes the Comptonization of soft photons in a hot plasma. The soft photon input spectrum is assumed to follow a Wien law. The soft photons experience inverse Compton scattering in an electron cloud whose temperature can range from a few keV up to 500 keV. This model includes relativistic effects and works for both the optically thin and thick regimes. \comptt\ has four parameters; the seed photon temperature for the assumed Wien spectrum (kT$_{\rm w}$), the electron temperature (kT$_{\rm e}$), the optical depth ($\tau$) and a parameter describing the geometry of the comptonizing cloud, either spherical or disk-like. In the following, a spherical geometry is assumed; however a disk-like geometry works equally well. The \comptt\ model works for spectra of intermediate electron temperature and optical depth, as is the case for \U. 

\begin{figure}[!t]
  \centerline{\includegraphics[width=6cm]{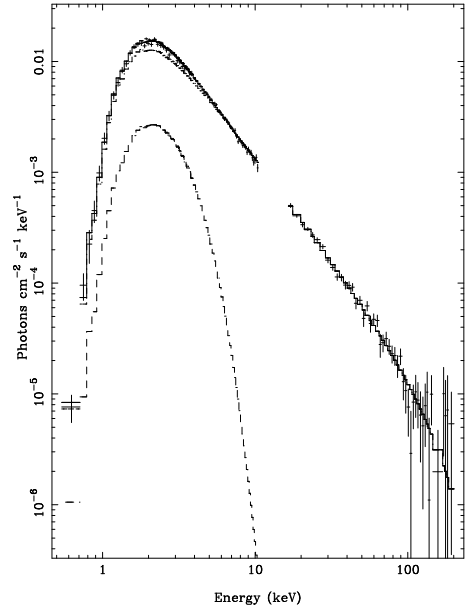}}
  \caption[]{The unfolded \sax\ spectrum of \U. This is the first report of a hard X-ray tail extending above 100 keV from the source. The soft component in dashed line is fitted with a single temperature blackbody. It contributes to $\sim 5$\% to the bolometric source luminosity. The high energy cutoff around 100 keV is statistically significant despite the relatively large error bars of the data.}
  \label{barret_fig2}
\end{figure}

Fitting the combined spectrum with an absorbed \comptt\ model reveals the presence of a soft excess below 5 keV. This soft excess can be fitted by either a blackbody or a multi-color disk blackbody. Similar two component models have been shown to describe the spectra of low luminosity LMXBs in their hard states \citep[see][]{guainazzi98,zand99,barret00, natalucci01,bo02}. The best fit spectral parameters and errors for the two models are listed in Table \ref{barret_table1}. As can be seen, the two models are statistically equivalent. The unfolded spectrum of \U\ together with its best fit is shown on Fig \ref{barret_fig2}, in which the soft component is fitted with a blackbody. During the \sax\ observation, the absorption corrected 1-200 keV luminosity was $\sim 2\times 10^{36}$ \ergs\ at 4 kpc.

\begin{table*}[!t]
\caption[]{ Best fit spectral results for \U~for two models: BB+C is the sum of a blackbody and \comptt, and C+DBB is the sum of a disk blackbody and \comptt. The fitted interstellar column density \nh~given in units of $10^{22}$ cm$^{-2}$. R$_{\rm BB}$ is the equivalent blackbody radius at the distance of 4.0 kpc. $ kT_{\rm in}$ and $R_{\rm in} \sqrt{\cos \theta}$ are the parameters  describing the disk blackbody model: the temperature at the inner disk and the projected inner disk radius computed at the assumed source distance of 4 kpc ($\theta$ is the unknown source inclination angle). {See text for the description of the \comptt\ parameters} The errors are quoted at the 90\% confidence level. F$_{\rm Soft}$ and F$_{\rm C}$ are the 1-200 keV bolometric flux of the soft and \comptt\ components respectively. F$_{\rm Tot}$ is the total bolometric flux. All fluxes are given in units of $10^{-9}$ \ergscm. Finally, the \chisq~and the number of degrees of freedom (d.o.f) are also listed.}
\begin{center}
\begin{tabular}{lcccccccccc}
Mod & N$_{\rm H}$ & kT$_{\rm BB}$ or kT$_{\rm in}$ & R$_{\rm BB}$ or R$_{\rm in}$ & kT$_{\rm w}$ & kT$_{\rm e}$ & $\tau$ & F$_{\rm Soft}$ & F$_{\rm C}$ & F$_{\rm Tot}$ &\chisq ~(d.o.f) \\
\noalign{\smallskip\hrule\smallskip}
{BB+C} &   $1.5^{+0.2}_{-0.2}$ & $0.64^{+0.04}_{-0.05}$ & $2.0^{+0.7}_{-0.3}$  & $0.34^{+0.06}_{-0.08}$ & $35.9^{+147.0}_{-7.0}$ & $2.9^{+0.6}_{-2.3}$ & 0.04 & 1.05 & 1.09 & 139.1 (130) \\
\noalign{\smallskip\smallskip}
{DBB+C} &   $1.6^{+0.2}_{-0.2}$ & $0.90^{+0.1}_{-0.1}$ & $1.0^{+0.3}_{-0.2}$  & $0.35^{+0.05}_{-0.07}$ & $36.0^{+78.0}_{-9.0}$ & $3.0^{+0.7}_{-1.8}$ & 0.06 & 1.01 & 1.07 & 134.4 (130) \\
\noalign{\smallskip\hrule\smallskip}
\end{tabular}
\end{center}
\label{barret_table1}
\end{table*}

As can be seen in Table \ref{barret_table1}, the electron temperature derived from \comptt\ has relatively large error bars. It is thus worth asking whether the high energy cutoff is statistically significant in the PDS spectrum. To test its significance, we have fitted the PDS data alone with both the \comptt~model and a power law. For the \comptt~model we have frozen the seed photon temperature to the value determined from the previous fit, because it cannot be determined from the PDS data alone. The power law fit is much worse with a \chisq~of 54.8 (45 d.o.f) as whereas it is only 43.7 (44 d.o.f) for the \comptt~model. This is because the power law cannot account for the high energy cutoff.  Very similar numbers are obtained if instead of the \comptt\ model, we used its approximation with a cutoff power law model. From an F-test, we conclude that the high energy cutoff is significant at more than a 99.8\% confidence level.


We have carefully searched for emission lines around 6 keV, as such lines have been reported from similar systems \citep[e.g.][]{barret00}. No emission lines were detected. A $2\sigma$ upper limit of $\sim 50$ eV was derived for the equivalent width of a  6.4 keV iron line of 0.2 keV width.

\subsection{\rxte}
\U~was simultaneously observed with \rxte~on April 20, 2000 between 18:56 and 22:49. The PCA instrument aboard \rxte~consists of a set of five identical Xenon proportional counter units (PCUs) covering the 2-60 keV range with a total area of about 6500 cm$^2$. The HEXTE instrument is made of two clusters (0 and 1) of four NaI(Tl)/CsI(Na) phoswich scintillation detectors \citep[see][ for a full description of RXTE]{bradt93}.

We have analyzed the \rxte~data using the latest version of HEASOFT (5.2) and the automated {\it rex} 0.3 script. During the observation, only PCU 0 and 2 were working nominally. Good time intervals were computed requesting these two PCUs to be ON, and using standard filtering criteria: earth elevation angle greater than 10 degrees, pointing offset less than 0.02 degrees, time since the peak of the last SAA passage greater than 30 minutes, and electron contamination less than 0.1. This corresponds to a net PCA exposure time of 6.8 kseconds. The PCA and HEXTE background subtracted light curves and spectra were then extracted. The source was detected with a steady count rate of $\sim 35$ counts s$^{-1}$ in PCU0 between 2.5 and 30 keV, and at a rate of $\sim 7$ counts s$^{-1}$ in HEXTE between 20 and 100 keV (cluster 0). No X-ray bursts were detected. We have checked that the combined PCA and HEXTE spectra could be fitted with the same models as the ones derived from the \sax~data. However the spectral parameters are obviously much less constrained with the \rxte~data, which lack statistics in the HEXTE range and spectral resolution in the PCA band pass.

\begin{figure}[!h]
  \centerline{\includegraphics[width=8cm]{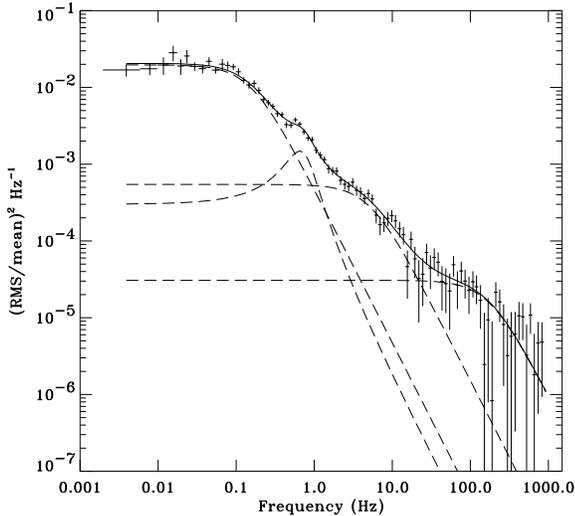}}
  \caption[]{The Fourier Power Density Spectrum of \U~computed between 3 and 30 keV from the RXTE/PCA data. It displays four distinct noise components; one of which extends above 200 Hz. All components were fitted with Lorentzians. The characteristic frequency of the first Lorentzian modeling the break and the frequency of the second one accounting for a low frequency quasi-periodic oscillation fit within the global correlation found by \citet{wijnands99}.}
  \label{barret_fig3}
\end{figure}

The scope of the \rxte~observation was rather to measure the rapid X-ray variability of the source during the \sax~observation. Within the good time interval found above, we have analyzed the {high time resolution data, recorded in the mode  E\_125us\_64M\_0\_1s, meaning that events are time-stamped with a 125-microsecond resolution in 64 PHA channel bands starting at channel 0, one buffer being readout every 1 second}. A power density spectrum was computed in the 0.015-2048 Hz range between 3 and 30 keV. The Poisson counting noise level was estimated above 1000 Hz and subtracted from the PDS. The PDS was rebinned using a logarithmic scheme, and expressed in terms of fractional RMS amplitude. It is shown in Fig \ref{barret_fig3}.

The PDS shown in figure \ref{barret_fig3} is very similar to those observed from low state LMXBs. It has been shown that these PDS can be well represented as  a superposition of up to 4 Lorentzians \citep[see e.g.][]{olive98,belloni02,vanstraaten02}. {These Lorentzians are either zero-centered to account for band limited noise (shot noise), or peaked to account for quasi-periodic oscillations. As stressed by \citet{belloni02}, the multi-Lorentzian approach gives a simple and universal phenomenological description of the PDS of a wide variety of sources, including accreting black holes and neutron stars}. Each Lorentzian is then described by its characteristic frequency, width (defined by the quality factor Q=$\nu / \Delta \nu$) and amplitude, usually expressed as a fractional RMS. 

The PDS of \U~can be represented by the sum of four Lorentzians, analytically defined in \citet{belloni02} \citep[see also][]{vanstraaten02,olive02}. {Three of them are zero-centered (Q=0), associated with aperiodic noise components, and one has a Q value of 1 to account for the presence of a short-lived oscillation at a frequency $\nu_{\rm LF}$=0.7 Hz}. The best fit parameters are given in Table \ref{barret_table2}, where each of the Lorentzians is named according to the definition given in \citet{belloni02}. Note that the source displays strong rapid X-ray variability as the total RMS is about 35\%, another common feature of low state LMXBs.

\begin{table}[!t]
\caption[]{Best fit spectral results for the power density spectrum of \U. The name of the Lorentzians follow the convention of \citet{belloni02}.}
\begin{center}
\begin{tabular}{cccc}
Lorentzian & Frequency & Q value & RMS (\%) \\
\noalign{\smallskip\hrule\smallskip}
$\nu_b$ & $0.15_{-0.1}^{+0.2}$ & Q=0 (fixed)  &$9.8_{-0.2}^{+0.2}$ \\
$\nu_{LF}$ & $0.73_{-0.05}^{+0.07}$ & $0.9_{-0.2}^{+0.2}$ & $3.9_{-0.3}^{+0.3}$ \\
$\nu_l$ & $5.3_{-0.7}^{+0.7}$ & Q=0 (fixed)  & $9.5_{-0.4}^{+0.4}$ \\
$\nu_u$ & $177.0_{-64.0}^{+110.0}$ & Q=0 (fixed) & $13.1_{-1.6}^{+1.8}$ \\
\noalign{\smallskip\hrule\smallskip}
\end{tabular}
\end{center}
\label{barret_table2}
\end{table}

\section{Discussion}
We have presented here the first broad band energy spectrum of \U, and its first detection at 100 keV. Thus \U~joins the list of the $\sim 20$ X-ray bursters presently detected at $\sim 100$ keV \citep[see the list in][]{disalvo02}. We have also reported on the first measurement of the power density spectrum of its rapid X-ray variability. 

\U~clearly belongs to the class of  low luminosity LMXBs displaying persistent and weakly variable emission \citep[other examples are SLX1735-269 and GS1826-234, see e.g.][]{barret00}. This is clearly illustrated by its long term ASM light curve (Fig. \ref{barret_fig1}). At the time of the \sax~and \rxte~observations, the bolometric luminosity of \U\ was about $\sim 2 \times 10^{36}$\ergs~(1-200 keV), assuming the source distance of 4 kpc derived by \citet{cocchi00}. This is very close to its mean luminosity. About half of this luminosity is radiated in X-rays between 1 and 20 keV.  Its 1-20 and 20-200 keV  luminosities fall at the bottom left of the so-called X-ray burster box \citep{barret00,disalvo02}.

The energy spectrum recorded by \sax~resembles those observed from these systems in their low state \citep[see][ for a review]{barret01,disalvo02}. It consists of the sum of a dominant hard comptonized component and a comparatively weaker soft component. The hard component is commonly thought to arise from a hot inner flow developing between the neutron star surface and a truncated accretion disk \citep[see e.g.][]{olive02}. The present data do not allow to tell whether the soft component is a single temperature blackbody or a multicolor disk blackbody, leaving two possible interpretations on its origin; either the neutron star surface or the truncated accretion disk. However, as it has been previously found in many other sources, the inner disk radius derived from a multicolor blackbody fit is very small, much smaller than any plausible neutron star radius \citep[e.g.][]{gierlinski02}. This has been interpreted as evidence against the disk origin for the soft component. So it more likely arises from  the neutron star, where only a fraction of its surface can be seen through the comptonizing cloud. This is consistent with the small  equivalent blackbody radius measured \citep[e.g.][]{gierlinski02}.

The X-ray variability of \U\ measured by \rxte~is also typical of low luminosity LMXBs. It shows several broad noise components, one of which extends above 200 Hz. The later feature is common to low state LMXBs and has been proposed as a way to distinguish black holes from neutron stars \citep{sunyaev00}. It has been shown that the frequency of the first Lorentzian ($\nu_b$ in Table 2) accounting for the flat top of the power spectrum and the frequency of the peaked feature immediately after the break ($\nu_{LF}$ in Table 2) are nicely correlated among a wide sample of sources, including black holes and neutron stars \citep{wijnands99,belloni02}. \U~is no exception and falls also on the correlation, {close to the data points of the low luminosity X-ray bursters 1E1724-3045 and GS1826-24} \citep[see Figure 11 in][]{belloni02}. {Thus, our data points strengthen this correlation which strongly suggests that the flat top noise and the low-frequency QPO are caused by the same unknown physical mechanism in all these sources \citep{wijnands99}.}

Finally, another correlation has been reported between $\nu_{LF}$ and $\nu_l$;  $\nu_l$ is then interpreted as the frequency of a lower kHz QPO \citep{psaltis99,belloni02}. This correlation, which extends over $\sim 3$ decades of frequency was a firm prediction of the relativistic precession model \citep[see e.g.][]{stella99}. In \U\, $\nu_l$ is low compared to the values observed from similar systems (generally $\nu_l \ge 10$Hz in most sources). It is thus significantly lower than the value expected from the above correlation. {Obviously, one cannot draw definite conclusions based on one single object, but it would be worth investigating other low luminosity systems to check whether the above correlation really breaks down at frequencies below $\sim 10$ Hz, as suggested by the present data. This would have important implications of the physical interpretation of these signals as relativistic timing features.}
\section{Conclusions}
The \sax~and \rxte~observations presented here have shown that \U~has spectral and timing properties typical of low luminosity neutron star low mass X-ray binaries; displaying both a hard X-ray tail and intense rapid X-ray variability. {\U\ enlarges the list of twenty X-ray bursters presently detected at 100 keV. Its broad band energy spectrum can be fitted by the sum of a dominant hard Comptonized component (electron temperature of $\sim 36$ keV and optical depth $\sim 3$) and a weak soft component. The latter component which can be fitted with a blackbody of $\sim 0.6$ keV and equivalent radius of $\sim 2$ km is likely to originate from the neutron star surface. As expected for a low luminosity system, the power density spectrum measured by RXTE is characterized by the presence of a $\sim 0.7$ Hz low frequency quasi-periodic oscillation together with three broad noise components, one of which extends above $\sim 200$ Hz.}

The location of the source in the galactic plane, its persistent intensity and hard spectrum mean that it will be repeteadly detected by the IBIS hard X-ray imaging instrument aboard INTEGRAL during its weekly scans \citep{lebrun01ibis}. This will provide insights on the long term spectral evolution and timing properties of its hard X-ray tail.

\begin{acknowledgements}
We are grateful to Matteo Guainazzi for stimulating discussions at the early stage of this project. The \sax\ satellite is a joint Italian-Dutch programme.  We thank the \sax\ and RXTE planners; Donatella Ricci and Evan Smith for making the simultaneous observations of \U~possible. This research has made use of data obtained from the High Energy Astrophysics Science Archive Research Center (HEASARC), provided by NASA's Goddard Space Flight Center.
\end{acknowledgements}

\end{document}